\documentclass{article}
\usepackage{ijcai25}

\usepackage{times}
\usepackage{soul}
\usepackage{url}
\usepackage[hidelinks]{hyperref}
\usepackage[utf8]{inputenc}
\usepackage[small]{caption}
\usepackage{graphicx}
\usepackage{amsmath}
\usepackage{amsthm}
\usepackage{booktabs}
\usepackage{algorithm}
\usepackage{algorithmic}
\usepackage[switch]{lineno}
\usepackage{amssymb}
\usepackage[round]{natbib}   

\title{QA-MDT: Quality-aware Masked Diffusion Transformer for
Enhanced \\ Music Generation}

\author{
Chang Li$^{1*}$\and
Ruoyu Wang$^{1*}$\and
Lijuan Liu$^1$\and
Jun Du$^{1\dagger}$\and
Yixuan Sun$^1$\and
Zilu Guo$^1$\and \\
Zhengrong Zhang$^1$\and
Yuan Jiang$^1$\and
Jianqing Gao$^2$\and
Feng Ma$^2$\\
\affiliations
$^1$University of Science \& Technology of China, Hefei, China\\
$^2$iFlytek AI Research, Hefei, China\\
\emails
\{lc\_lca, wangruoyu\}@mail.ustc.edu.cn,
jundu@ustc.edu.cn
}

\begin{document}

\maketitle

\begin{abstract}
Text-to-music (TTM) generation, which converts textual descriptions into audio, opens up innovative avenues for multimedia creation.
Achieving high quality and diversity in this process demands extensive, high-quality data, which are often scarce in available datasets. Most open-source datasets frequently suffer from issues like low-quality waveforms and low text-audio consistency, hindering the advancement of music generation models.
To address these challenges, we propose a novel quality-aware training paradigm for generating high-quality, high-musicality music from large-scale, quality-imbalanced datasets. Additionally, by leveraging unique properties in the latent space of musical signals, we adapt and implement a masked diffusion transformer (MDT) model for the TTM task, showcasing its capacity for quality control and enhanced musicality. Furthermore, we introduce a three-stage caption refinement approach to address low-quality captions' issue. Experiments show state-of-the-art (SOTA) performance on benchmark datasets including MusicCaps and the Song-Describer Dataset with both objective and subjective metrics.
Demo audio samples are available at https://qa-mdt.github.io/, code and pretrained checkpoints are open-sourced at https://github.com/ivcylc/OpenMusic/.
\end{abstract}
%

\section{Introduction}
\begin{figure}
    \centering
    \includegraphics[width=1\linewidth]{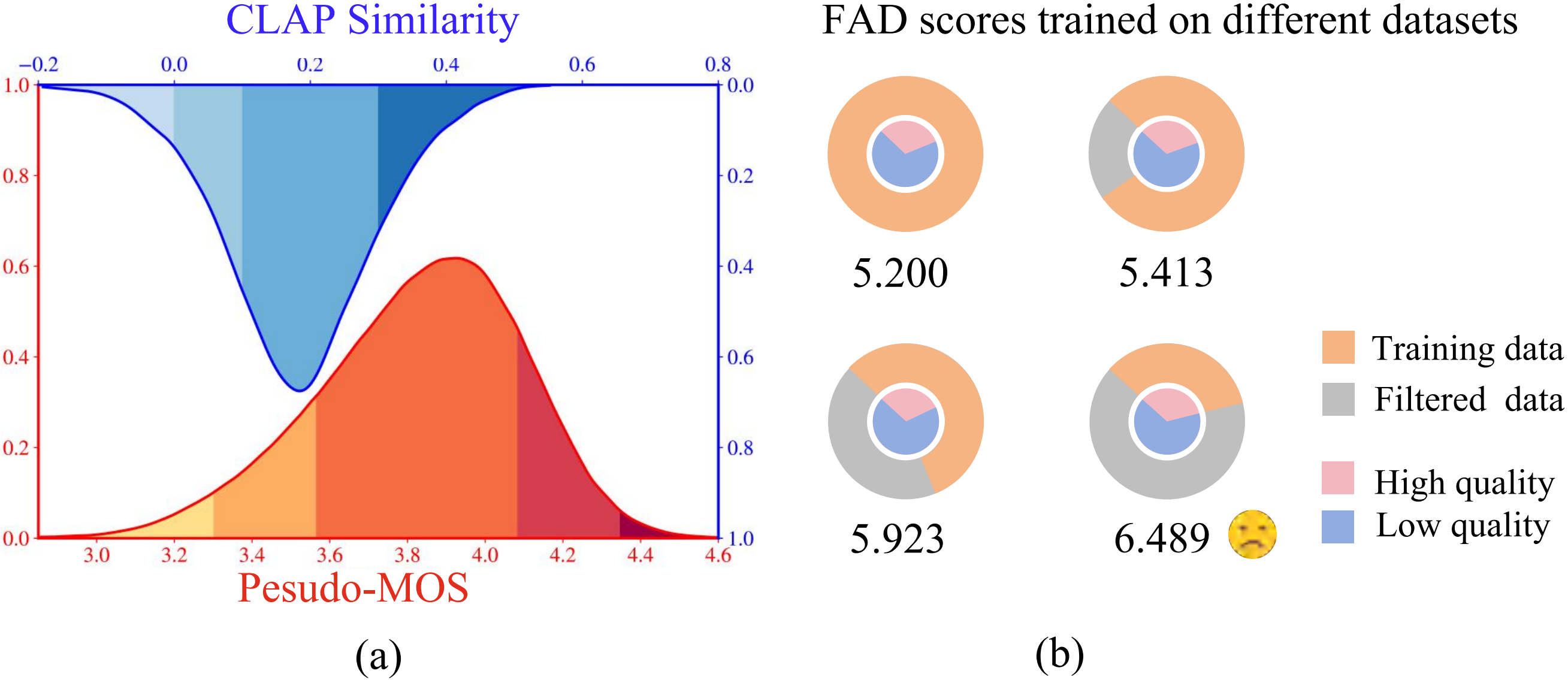}
    \caption{\textbf{(a)} The distribution curves of CLAP similarity and pseudo-MOS for large-scale open-source music databases AudioSet~\citep{defferrard2016fma} and FMA~\citep{defferrard2016fma}, where darker areas represent higher text-audio consistency or audio quality. \textbf{(b)} Frechet Audio Distance (FAD)~\citep{kilgour2018fr} scores on the MusicCaps test set obtained from models trained for 50K steps on AudioSet and FMA, using filter ratios of 0\%, 33\%, 66\%, and 100\% of low-quality data. Here, low-quality data is determined by a Pseudo-MOS score~\citep{ragano2023audio} of less than 4.0. It can be inferred that performance consistently worsens with larger filter ratios.}
    \label{fig:enter-label}
\end{figure}
Text-to-music (TTM) generation aims to transform textual descriptions of emotions, style, instruments, rhythm, and other aspects into corresponding music segments, providing new expressive forms and innovative tools for multimedia creation. According to scaling law principles~\citep{peebles2023scalable, li2024scalability}, effective generative models require a large volume of training data. However, unlike image generation tasks~\citep{chen2024pixart, rombach2021highresolution}, acquiring high-quality music data often presents greater challenges, primarily due to copyright issues and the need for professional hardware to capture high-quality music. These factors make building a high-performance TTM model particularly difficult.

In the TTM field, high-quality music signals is scarce. This prevalent issue of low-quality data, highlighted in Figure~\ref{fig:enter-label} (a), manifests in two primary challenges. Firstly, most available music signals often suffer from distortion due to noise, low recording quality, or outdated recordings, resulting in diminished generated quality, as measured by pseudo-MOS scores from quality assessment models~\citep{ragano2023audio}. Secondly, there is a weak correlation between music signals and captions, characterized by missing, weak, or incorrect captions, leading to low text-audio similarity, which can be indicated by CLAP scores~\citep{laionclap2023}. These challenges, especially the \textbf{inherent quality in the music signal itself}, significantly hinder the training of high-performance music generation models, resulting in poor rhythm, noise, and inconsistencies with textual control conditions in the generated audio. Additionally, as shown in Figure~\ref{fig:enter-label} (b), directly filtering low-quality music, which robustly reduces the dataset size, leads to a consistent decline in model performance. Therefore, finding an effective training strategy for large-scale datasets with low-quality waveforms, mismatches, and missing labels has become an urgent challenge.

In this paper, we introduce a novel quality-aware masked diffusion transformer (QA-MDT) to enhance music generation, aiming to tackle the aforementioned problems while making further improvements through architectural studies. We made efforts on effectively leveraging extensive open-source music databases, which often contain data of varying quality and style, to produce high-quality, diverse and high text-audio consistency music. \textbf{For music quality enhancement}, we innovatively inject music quality into the denoising stage with multiple granularities to foster quality awareness during training, while high-quality music can be obtained by setting a quality threshold during inference. \textbf{Regarding the modeling architecture}, in preliminary experiments, we found that the Diffusion Transformer (DiT) framework, which has been successful in the image domain~\citep{peebles2023scalable}, is not directly effective for modeling music spectrograms. However, injecting a masking strategy significantly enhances the spatial correlation of the music spectrum and further accelerates convergence. Additionally, we utilize large language models (LLMs) and the CLAP model to synchronize music signals with captions, thereby \textbf{enhancing text-audio correlation} in extensive music datasets. Our ablation studies on public datasets confirm the effectiveness of our methodology, with the final model surpassing previous works in both objective and subjective measures.
In summary, we focus on developing better training strategies and network architectures to enhance the quality and aesthetic of music generation. At the same time, we address the long-standing issue of text-audio consistency in the field of TTM, which can be listed as:
\begin{itemize}
    \item We propose a quality-aware training paradigm that enables the model to perceive the quality of the dataset during training, thereby achieving superior music generation in terms of both musicality and audio quality.
    \item We innovatively introduced the Masked Diffusion Transformer to music signals, demonstrating its unique efficacy in modeling music latent space and its capability in perceiving quality control, thereby further improving both the generated quality and musicality.
    \item We address the issue of low text-audio correlation in large-scale music datasets for TTM, effectively improving text alignment and generative diversity.
\end{itemize}

\section{Related Work}\label{sec:rel}

\paragraph{Text to music generation.}
Text-to-music generation aims to create music clips that correspond to input descriptive or summary text. Previous efforts have utilized either language models (LMs) or diffusion models (DMs) to model quantized waveform representations or spectral features. Models like MusicLM~\citep{agostinelli2023musiclm}, MusicGen~\citep{copet2024simple}, MeLoDy~\citep{lam2024efficient}, and Jen-1~\citep{li2024jen} leverage LMs and DMs on residual codebooks obtained via quantization-based codecs~\citep{zeghidour2021soundstream,defossez2022high}. Moûsai~\citep{schneider2023mo}, Noise2Music~\citep{huang2023noise2music}, Riffusion~\citep{forsgren_martiros_2022}, AudioLDM 2~\citep{liu2023audioldm2}, and Stable Audio~\citep{evans2024fast} use U-Net-related diffusion to model mel-spectrograms or latent representations obtained through compression networks. Although some approaches attempt to guide the model towards generating high-quality content by setting negative prompts like ``low quality''~\citep{liu2023audioldm2, chen2024musicldm}, few explicitly inject quality information during training. This results in the model's inability to effectively perceive and control content quality.

\paragraph{Transformer based diffusion models.}
Traditional diffusion models typically use U-Net as the backbone, where the inductive biases of CNNs do not effectively model the spatial correlations of signals and are insensitive to scaling laws~\citep{li2024scalability}. However, transformer-based diffusion models (DiT) \citep{peebles2023scalable} have effectively addressed these issues. This advantage is particularly evident in fields such as video generation~\citep{brooksvideo}, image generation~\citep{peebles2023scalable,chen2024pixart,bao2022all}, and speech generation~\citep{liu2023vit}. To expedite training and foster inter-domain learning of correlations, the masking strategy has proven effective, yielding SOTA class-conditioned performances on ImageNet~\citep{gao2023masked}. Additionally, a simpler architecture~\citep{zheng2023fast} incorporating reconstruction losses and unmasked fine-tuning further enhances model training speed.
However, these models have not yet been verified for text-controlled music generation on large-scale music datasets, and their adaptability with additional control information remains an open question. Make-an-audio 2~\citep{huang2023make} and, more recently, Stable Audio 2~\citep{123evans2024long}, have explored the DiT architecture for audio and sound generation. However, their approach models latent tokens by segmenting only along the time dimension to control and extend generation duration. In contrast, our focus is on finer segmentation within the latent space across both time and frequency, aiming for more precise modeling of music signals.

\paragraph{Quality enhancement in audio domain.}
Previous research has made efforts to improve the quality of generated audio, particularly in two key areas: waveform fidelity and the consistency between input text and generated content. Waveform quality can be compromised by issues like aliasing from low sampling rates and limited expressiveness due to monophonic representations, while models like MusicGen~\citep{copet2024simple} and Stable Audio~\citep{evans2024fast, 123evans2024long}, which directly model 32k and 44.1k stereo audio, have significantly enhanced perceptual quality. Despite higher sampling rates and channels, the quality of audio in training datasets remains inconsistent, often suffering from noise, dullness, and a lack of rhythm or structure. These problems, often reflected by the Mean Opinion Score (MOS), are rarely addressed. In terms of text-audio consistency, Make-an-audio 2~\citep{huang2023make} and WavCaps~\citep{mei2024wavcaps} have employed ChatGPT-assisted data augmentation to improve temporal relationships and accuracy in audio effect generation. Although studies like Music-llama~\citep{liu2024music} and LP-musiccaps~\citep{doh2023lp} have introduced captioning approaches for music, few have explored the augmentation and utilization of synthetic data in large-scale music generation tasks. Additionally, \citet{gui2024adapting} overhaul FAD by introducing a sample-size–extrapolated score, curated high-quality reference sets, and music-aware embeddings, producing scores that align much more closely with human judgments of music quality.

\section{Preliminary}\label{sec:rel}

\paragraph{Latent diffusion model.}
Direct application of DMs to cope with distributions of raw signals incurs significant computational overhead~\citep{ho2020denoising, song2020denoising}. Conversely, studies~\citep{liu2023audioldm, liu2023audioldm2} apply them in a latent space with fewer dimensions. 
The latent representation \( z_0 \) is the ultimate prediction target for DMs, which involve two key processes: diffusion and reverse processes. In the diffusion process, Gaussian noise is incrementally added to the original representation at each time step \( t \), described by \( z_{t+1} = \sqrt{1-\beta_t} z_t + \sqrt{\beta_t} \epsilon \), where \( \epsilon \) is drawn from a standard normal distribution \( \mathcal{N}(0, I) \), and \( \beta_t \) is gradually adapted based on a preset schedule to progressively introduce noise into the state \( z_t \).
The cost function~\citep{ho2020denoising,liu2023audioldm} is formalized as
\(
\arg\min_{\theta}  \mathbb{E}_{{(z_0, y), \epsilon}} \left[ \left\| \epsilon - D_\theta\left(\sqrt{\alpha_t}z_0 + \sqrt{1-\alpha_t}\epsilon, t, y\right) \right\|^2 \right] 
\).
where \( D_\theta \), the denoising model, strives to estimate the Gaussian noise \( \epsilon \), conditioned on the latent state \( z_t \), the time step \( t \), the conditional embedding \( y \), and where \(\alpha_t\) represents a predefined monotonically increasing function.
{In the reverse process, we obtain \( z_{t-1} \) via the recursive equation:
\(
z_{t-1} = \frac{1}{\sqrt{1- \beta_t}}\left( z_{t} - \frac{\beta_t}{\sqrt{1-\alpha_t}}\epsilon_\theta \right)  + \sqrt{\frac{1-\alpha_{t-1}}{1-\alpha_t}\beta_t}  \epsilon\),
 where \(\epsilon_\theta\) represents the estimated Gaussian noise.}

\paragraph{Classifier-free guidance.}
Classifier-free guidance (CFG), introduced by~\citep{ho2020denoising}, increases the versatility and flexible control ability of DMs by both considering conditional and unconditional generation. Typically, a diffusion model generates content based on specific control signals \( y \) within its denoising function \( D_{\theta}(z_t, t, y) \). CFG enhances this mechanism by incorporating an unconditional mode \( D_{\theta}(z_t, t, \emptyset) \), where \( \emptyset \) symbolizes the absence of specific control signals. The CFG-enhanced denoising function is then expressed as \( D_{\theta}^{\text{CFG}}(z_t, t, y) = D_{\theta}(z_t, t, y) + w(D_{\theta}(z_t, t, y) - D_{\theta}(z_t, t, \emptyset)) \), where \( w \geq 1 \) denotes the guidance scale.
During training, the model substitutes \( y \) with \( \emptyset \) at a constant probability \( p_{\text{uncond}} \). In inference, \( \emptyset \) might be replaced by a negative prompt like ``low quality'' to prevent the model from producing such attributes~\citep{liu2023audioldm2}.

\section{Method}\label{sec:method}

\begin{figure*}[t]
    \centering
    \includegraphics[width=\textwidth]{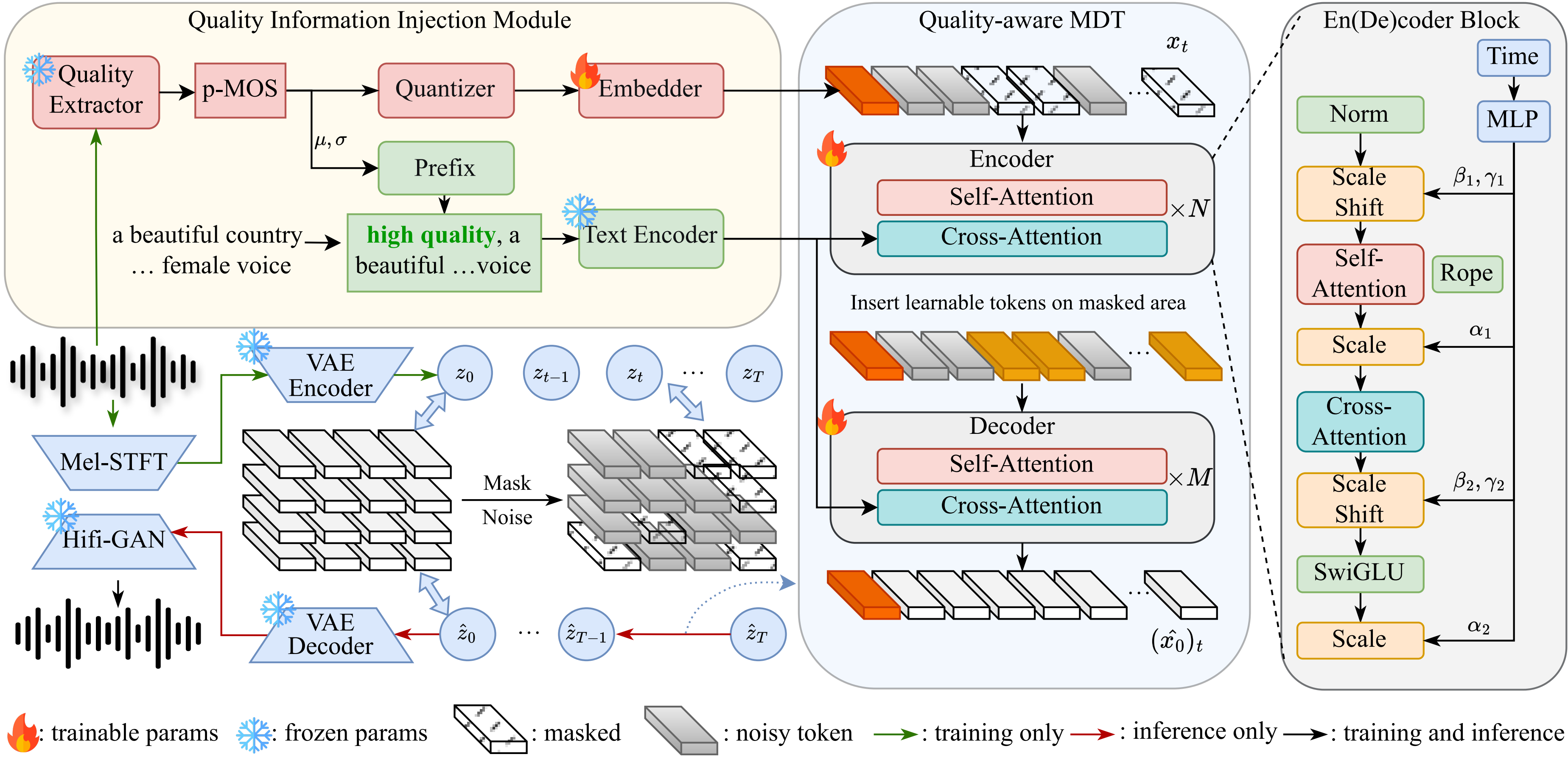}
    \caption{Pipeline of proposed quality-aware masked diffusion transformer for music generation.}
    \label{fig:kuangjia}
\end{figure*}

\subsection{Quality Information Injection}
\label{quality_informaton_injection}
At the heart of our work lies the implementation of a pseudo-MOS scoring model~\citep{ragano2023audio} to meticulously assign music quality to quality prefixes and quality tokens. 

We define our training set as \(\mathcal{D}_o = \{(M_i, T^{o}_i) \mid i = 1, 2, \dots, N_D\}\), where each \(M_i\) represents a music signal and \(T^{o}_i\) is the corresponding original textual description.
To optimize model learning from datasets with diverse audio quality and minimize the impact of low-quality audio, we initially assign \textit{p}-MOS scores to each music track using a model fine-tuned with wav2vec 2.0~\citep{baevski2020wav2vec} on a dataset of vinyl recordings for audio quality assessment, and achieve the corresponding \textit{p}-MOS set \( S = \{s_1, s_2, \ldots, s_{N_D}\} \). These scores facilitate dual-perspective quality control for enhanced granularity and precision. 

First, We analyze this \textit{p}-MOS set \(S\) to identify a negative skew normal distribution with mean \( \mu \) and variance \( \sigma^2 \). We define text prefixes based on \( s \) as follows: prepend ``low quality'' if \( s < \mu - 2\sigma \), ``medium quality'' if \( \mu - \sigma \leq m \leq \mu + \sigma \), and ``high quality'' if \( s > \mu + 2\sigma \). This information is prepended before processing through the text encoder with cross-attention, enabling the initial separation of quality-related information.

To achieve a more precise awareness and control of waveform quality, we synergize the role of text control with quality embedding. We observed that the distribution of \textit{p}-MOS in the dataset is approximately normal, which can be shown in Figure \ref{fig:enter-label}, allowing us to use the Empirical Rule to segment the data accordingly. Specifically, we define the quantization function \( Q: [0, 5] \to \{1, 2, 3, 4, 5\} \) to map the \textit{p}-MOS scores to discrete levels based on the distance from the mean \( \mu \) in terms of standard deviation \( \sigma \):
\begin{equation}
Q(s) = \left\lfloor \frac{s - (\mu - 2\sigma)}{\sigma} \right\rfloor + r
\end{equation}

where \(r=2\) 
for \(s>\mu\), otherwise, \(r=1\). 
Subsequently, \( Q(s) \) is mapped to a \( d \)-dimensional quality vector embedding using the embedding function \( E \), such that
\begin{equation}
q_{\text{vq}}(s) = E(Q(s)) \in \mathbb{R}^d,
\end{equation}
This process provides finer granularity of control within the following model and facilitates the ability of interpolative quality control during inference, enabling precise adjustments in \( \mathbb{R}^d \). In later stages, the quality embedding is treated as a token on par with every latent audio patch, participating in the attention computations to enable interaction.

\subsection{Quality-aware Masked Diffusion Transformer}
\label{model_architecture}
In a general patchify phase with patch size \(p_f \times p_l\) and overlap size \(o_f \times o_l\), patchified token sequence \(X=\{x_1, x_2, \dots, x_{P}\} \subset \mathbb{R}^{p_f \times p_l}\) are obtained through spliting the music latent space \(\mathcal{M}_{spec} \in \mathbb{R}^{F \times L}\), as described in Section ~\ref{sec:vae}. The total number of patches \(P\) is given by:
\begin{equation}
P = \left\lceil \frac{L - p_l}{p_l - o_l} + 1 \right\rceil \times \left\lceil \frac{F - p_f}{p_f - o_f} + 1 \right\rceil
\end{equation}

A 2D-Rope position embedding~\citep{su2024roformer} is added to each patch for better modeling of relative position relationship while a binary mask \(\boldsymbol{m} \in \{0, 1\}^{P}\) is applied during the training stage, with a variable mask ratio \(\gamma\). This results in a subset of \(\lfloor \gamma P \rfloor\) patches being masked that \(\sum_{i=1}^{P_N} m_i = \lfloor \gamma P \rfloor\), leaving \(P - \lfloor \gamma P \rfloor\) patches unmasked. The subset of masked tokens is invisible in the encoder stage and replaced with trainable mask tokens in the decoder stage following the same strategy utilized in AudioMAE~\citep{huang2022masked} and MDT~\citep{gao2023masked}.

The transformer we use consists of \(N\)  encoder blocks, \(M\) decoder blocks, and an intermediate layer to replace the masked part with trainable parameters. We treat the embedding of the quantized \textit{p}-MOS score as a prefix token, concatenated with each stage's music tokens. Let \({X}^k = [{x}^k_1, {x}^k_2, \ldots, {x}^k_{P}] \in \mathbb{R}^{P \times d}\) represent the output of \(k\)-th  encoder or decoder block, where the initial input of the encoder \(X^0=z_t=\alpha_t{z}_0 + \sqrt{1-\alpha_t}\epsilon\), and the final decoder block estimate  \(X^{N+M}=z_0 = [{x}_1, {x}_2, \ldots, {x}_{P}]\).
For \( k < N \), indicating the encoder blocks, the sequence transformation focuses only on unmasked tokens:
\begin{equation}
[q_\text{vq}^{k+1}; {X}^{k+1}] = \text{ Encoder}^k\left(\left[q_\text{vq}; {X}^k \odot (\mathbf{1} - \boldsymbol{m})\right]\right),
\end{equation}
where \(\boldsymbol{m} \in \{0, 1\}^P\) is the mask vector, with \(1\) indicating masked positions and \(0\) for visible tokens.

For \(N < k < N + M \), indicating the decoder blocks, the full sequence including both unmasked tokens and learnable masked tokens is considered:
\begin{equation}
[q_\text{vq}^{k+1}; {X}^{k+1}] = \text{Decoder}^k \left(\left[q_\text{vq}; {X}^k\right]\right),
\end{equation}
where the previously masked tokens are now subject to prediction and refinement.
In the decoding phase, the portions that were masked are gradually predicted, and throughout this entire phase, the quality token \(q_\text{vq}(s)\) is progressively infused and optimized. Subsequently, the split patches are unpatchified while the overlapped area is averaged to reconstruct the output noise and every token contributes to calculating the final loss:
\begin{equation}
\small
\mathcal{L}(\theta) = \mathbb{E}_{(z_0,
q_{vq}, y),\epsilon} \left[ \left\|\epsilon - D_{\theta} \left( \sqrt{\alpha_t} z_0
+ \sqrt{1-\alpha_t} \epsilon, t, q_{vq}, {y} 
 \right) \right\|^2 \right]
\end{equation}
In the inference stage, the model can be guided to generate high-quality music through modified CFG: 
\begin{equation}
\begin{aligned}
D_{\theta}^{\text{High}}(z_t, t, q_{vq}^{\text{high}}, y) &= D_{\theta}(z_t, t, q_{vq}^{\text{high}}, y) + \\
& w\left(D_{\theta}(z_t, t, q_{vq}^{\text{high}}, y) - D_{\theta}(z_t, t, q_{vq}^{\text{low}}, \emptyset)\right)
\end{aligned}
\end{equation}
Here \(q_{vq}^\text{high}\) and \(q_{vq}^\text{low}\) indicate quantified \textit{p}-MOS for guiding the model in a balance between generation quality and diversity.
After obtaining the sampled latent \(\hat{z_0}\) with DDIM sampler, we can finally generate the music using the VAE decoder.
\subsection{Music Caption Refinement}
\label{sec:audio_tokenization}
\looseness=-1
We divided the caption refinement stage into three steps including text information enriching with music caption model \(\mathcal{F}_\text{cap}\), caption adjustment with CLAP cosine similarity function \(\mathcal{S}\) and caption diversity extension with LLMs which we denoted as \({\mathcal{F}}_{\text{llm}}\).

Initially, pretrained music caption model~\citep{doh2023lp} is employed to re-annotate each music signal $M_i$ to $T^{g}_i$,  shown as 
$
\mathcal{D}_{g} = \{(M_i, T^{g}_i) \mid T^{g}_i = \mathcal{F}_\text{cap}(M_i), i = 1, 2, \dots, N\}
$.
CLAP text-audio similarity is applied to filter \(\mathcal{D}_{g}\) with a threshold of \(\rho_1\), resulting in 
\begin{equation}
\mathcal{D}^{\text{filter}}_{g} = \{(M_i, T^{g}_i) \mid \mathcal{S}(T^{g}_i, M_i) > \rho_1\}
\end{equation}
In this context, we meticulously filter out generated captions that do not correspond with their respective audio files. This misalignment may be attributed to inaccuracies within the captioner's insufficient training. For the filtered data pairs, we opt to retain the use of the original captions.

To ensure that valuable information from the original captions is not overlooked when using only the generated captions, we adapt a fusing stage to combine the original caption and generated pseudo prompt. Firstly, we need to filter out original captions that is useless or inaccurate, formulated as:
\begin{equation}
\mathcal{D}^{\text{filter}}_{o} = \{(M_i, T^{o}_i) \mid \mathcal{S}(T^{o}_i, M_i) > \rho_2\}.
\end{equation}
The issue can stem from the original data being improperly labeled with terms such as 'speech, car' from datasets like AudioSet~\citep{gemmeke2017audio} and also may be because of desperately missing of the original labels.

Finally, only the original caption that suffers low CLAP text similarity score should be merged with the generated ones, for redundant, repetitive parts result in long and verbose final captions. Thus, we set the threshold to \(\rho_3\) and merge them by LLMs to \(T_{\text{fusion}} = \mathcal{F}_{\text{llm}}(T^{o}, T^{g})\):
\begin{equation}
\begin{aligned}
\mathcal{D}_{\text{merge}} = \big\{ (M_i, T_{\text{fusion}}) \mid \mathcal{S}(T^{o}, T^{g}) < \rho_3, \\
\hspace*{2.5em}(M_i, T^{o}) \in \mathcal{D}^{\text{filter}}_{o}, (M_i, T^{g}) \in \mathcal{D}^{\text{filter}}_{g} \big\}.
\end{aligned}
\end{equation}


\section{Experimental Setup}\label{sec:exp}

\subsection{Datasets}\label{sec:datasets}
For training, we used the following databases for our training: AudioSet Music Subset (ASM)~\citep{gemmeke2017audio}, MagnaTagTune (MTT)~\citep{law2009evaluation}, Million Song Dataset (MSD)~\citep{bertin2011million}, Free Music Archive (FMA)~\citep{defferrard2016fma}, and an additional dataset\footnote{We use 55k music tracks from https://pixabay.com, which is a large scale copyright free dataset.}. Each track in these databases was clipped to 10-second segments and sampled at 16kHz to ensure uniformity across the dataset. The final training set was developed through a process of caption refinement, as detailed in Section~\ref{sec:audio_tokenization}. Finally, we got our training set totaling 12.5k hours of diverse music data. The specific composition of these datasets is further elaborated in the Appendix. 
For evaluation, we test our model on the widely used MusicCaps benchmark~\citep{agostinelli2023musiclm} and the Song-Describer-Dataset~\citep{manco2023thesong}. 
MusicCaps consists of 5.5K 10.24-second clips sourced from YouTube, each accompanied by high-quality music descriptions provided by ten musicians. The Song-Describer Dataset is made up of 706 licensed high quality music recordings.



\subsection{Models and Hyperparameters}\label{sec:hyperparams}

\paragraph{Audio compression.} 
\label{sec:vae}

Each 10.24-second audio clip, sampled at 16 kHz, is initially transformed into a \(64 \times 1024\) mel-spectrogram with mel-bins of 64, hop-length of 160 and window length of 1024. Subsequently, this spectrogram is compressed into a \(16 \times 128\) latent representation \(\mathcal{M}_{spec}\) using a Variational Autoencoder (VAE) pretrained with AudioLDM 2~\citep{liu2023audioldm2} with series of quantization loss and adversarial loss. We use pretrained Hifi-GAN~\citep{kong2020hifi} vocoder to reconstruct the waveform from the generated mel-spectrogram.

\paragraph{Caption processing and conditioning.} 
We utilize the LP-MusicCaps~\citep{doh2023lp} caption model for ASM, FMA, and subsets of MTT and MSD that have weak or no captions.
We use the official checkpoint from LAION-CLAP~\citep{laionclap2023}\footnote{\texttt{music\_speech\_audioset\_epoch\_15\_esc\_89.98.pt} } for text-to-text and text-to-audio similarity calculations.
Based on small scale subjective experienment, thresholds are set at \(\rho_1 = \rho_2 = 0.1 \) to ensure any generated text or original caption not aligned well with the corresponding waveform is filtered out. Additionally, after filtering, generated text that fall below a threshold of \(\rho_3 = 0.25 \) are merged with original tags with the prompt: \textit{Merge this music caption ``generated caption'' with the ground truth tags ``original tags'', and do not add any imaginary elements.}. We use FLAN-T5-large~\citep{peebles2023scalable} as text encoder for all models.

\paragraph{Diffusion backbone.} We train our diffusion model with three backbones for comparison: U-Net~\citep{ronneberger2015u} based at 1.0B parameters; our proposed Quality-aware Masked Diffusion Transformer (QA-MDT) with \(N=20\) encoder layers and \( M=8 \) decoder layers at 675M. We study the impact of the patch size and overlap size in the Appendix, and apply a patch size of \(1 \times 4\) without overlap for the training of our final model.
We train on 10.24-second audio crops sampled at random from the full track, maintaining a total batch size of 64, learning rate of 8e-5, and a condition drop of 0.1 during training. The final model was trained for a total of 38.5k steps. During inference, we use Denoising Diffusion Implicit Models (DDIM)~\citep{song2020denoising} with 200 steps and a guidance scale of 3.5, consistent with AudioLDM~\citep{liu2023audioldm}.
\label{sec:res}
We begin by presenting our approach to refining captioning, which includes the capability for quality awareness, transitioning from text-level control to token-level control. Finally, we compare proposed model with previous works
subjectively and objectively.
\subsection{Evaluation Metrics}
We evaluate the proposed method using objective metrics, including the Fréchet Audio Distance (FAD)~\citep{kilgour2018fr}, Kullback-Leibler Divergence (KL), Inception Score (IS)\footnote{We \textbf{strictly follow} the comparison method and evalution code in AudioLDM2 and ensure that the indicators in the paper \textbf{follow consistent sampling rates and durations for fair comparison}. All above metrics are computed using the \texttt{audioldm\_eval} library~\citep{liu2023audioldm}, ensuring standardized evaluation.}.
We also utilize pseudo-MOS scoring model~\citep{ragano2023audio} to estimate generation quality, with more accurate assessments derived from subjective metrics.


\section{Results}

\subsection{Quality Awareness}
\begin{figure*}[h]
    \centering
    \includegraphics[width=1\textwidth]{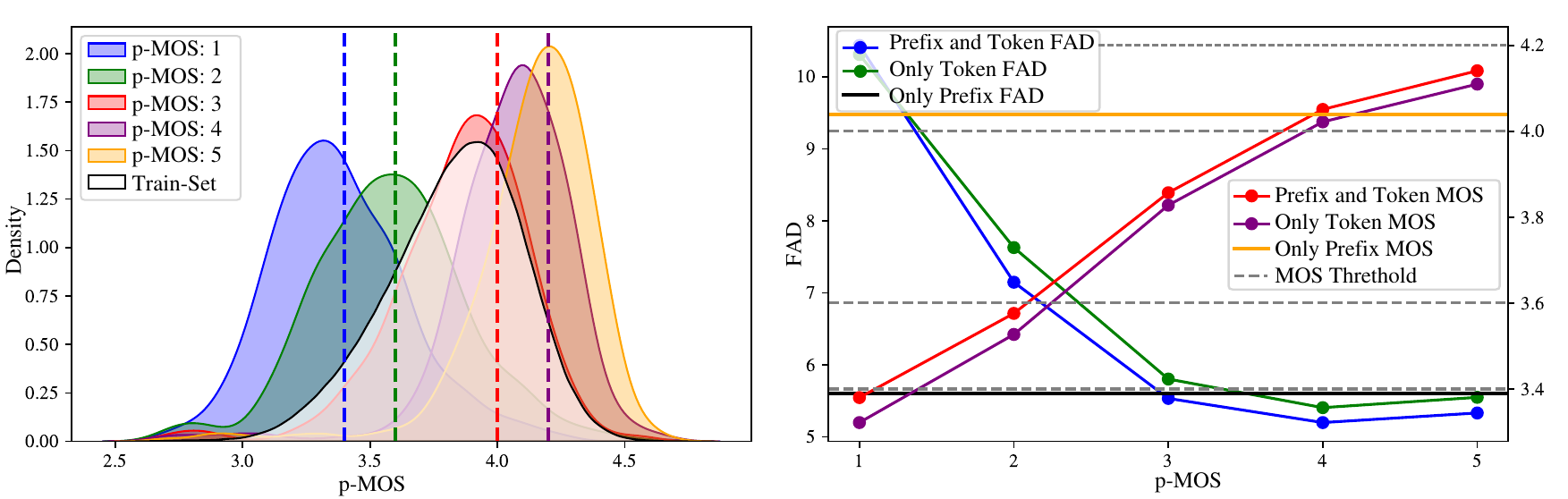}
    \caption{
\textbf{(Left)} Five \textit{p}-MOS distribution curves are obtained by concurrently using text quality prefixes and quality tokens as controls on the MTT-FS, with quantized MOS levels ranging from 1 to 5 serving as control constraint inferences. The distribution of the training set is normalized by each sample's duration, colored lines represent thresholds of quantized \textit{p}-MOS tokens during training. \textbf{(Right)} The effect of using quality text prefixes during training is shown, showcasing testing results on FAD and \textit{p}-MOS, while gray lines for quantized \textit{p}-MOS threshold.}
    \label{fig:2g}
\end{figure*}
\begin{figure}[ht]
    \centering
    \includegraphics[width=1\linewidth]{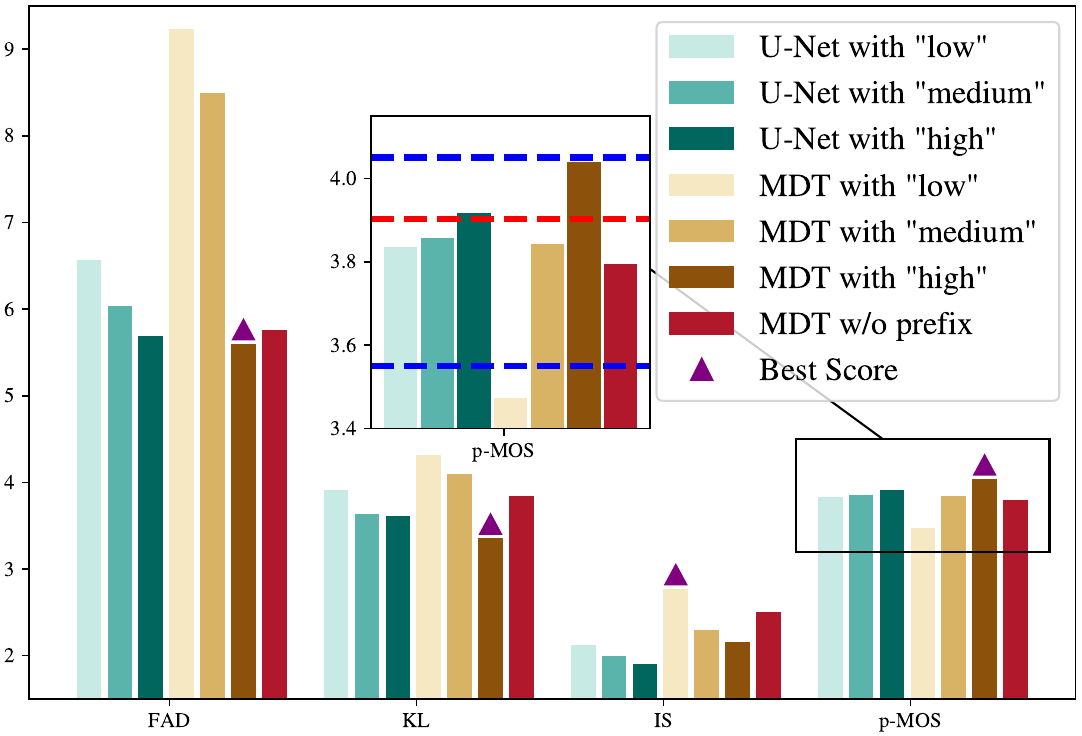}
    \caption{Comparison of model performance under different quality prefixes on MTT-FS, while the blue dashed line represents the threshold set during training to distinguish the three quality prefix levels, and the red one represents the test set average \textit{p}-MOS value.}
    \label{fig:pattern_abl}
\end{figure} 
This subsection explores the effects and interactions of model control over quality tokens and quality text prefixes during the training phase, as well as their comparative effects across different models. In our previous MTT dataset of 1,000 test pairs, we filtered out pairs labeled with \textit{low quality} or \textit{quality is poor} to avoid confusion when applying quality prefixes, resulting in a new subset of 519 entries, which we refer to as the MTT Filter Set (MTT-FS).
Figure~\ref{fig:pattern_abl} illustrates the impact of different quality prefixes during inference when quality is used as a text prefix during training for U-Net and MDT-based backbones. It was observed that U-Net, when inferred with different quality prefixes, showed only minor changes in \textit{p}-MOS scores and did not adhere to the threshold set during training. In contrast, MDT demonstrated better learning of quality information from prefixes, achieving \textit{p}-MOS scores significantly higher than those of U-Net and the test set.
Additionally, by decoupling quality information from the training set, we achieved better FAD (5.602 vs 5.757) and higher \textit{p}-MOS (4.039 vs 3.796) compared to training and inference without quality text prefixes. Given that quality tokens are specifically designed for the Transformer architecture, Figure~\ref{fig:2g} (left) shows the controlled outcomes when different quality tokens are used after integrating quantified quality as a token during training. Remarkably, using quality tokens alone provided more precise and accurate \textit{p}-MOS score control.
In our ablation study, we compared the effects of using only text prefixes against combining both approaches. As shown in Figure~\ref{fig:2g} (right), as the quantized control level gradually increased, the model steadily improved in \textit{p}-MOS scores, which represent the quality of generation. Concurrently, FAD and KL also progressively optimized until a turning point at level 4, where a higher average \textit{p}-MOS was achieved than when solely using prefixes. This turning point may be due to the scarcity of examples with quality level 5 in the dataset. Moreover, by combining two types of quality information injection, the refined decoupling and interaction allowed the model to more accurately perceive audio data quality features during training, leading to significant reductions in FAD and KL compared to using only one of them.

We also compare our approach with the traditional ``negative prompt'' strategy~\citep{liu2023audioldm2} in Appendix, highlighting our approach's significant improvement in quality and reduction in FAD.
\subsection{Impact of Music Caption Refinement}
\begin{table}[ht]
  \centering
  \setlength{\tabcolsep}{4pt} 
  \label{tab:main-results}
  
  
  \begin{tabular}{lccccccc}
    \toprule
    & \multicolumn{3}{c}{U-Net based} & \multicolumn{3}{c}{MDT based} \\
    \cmidrule(lr){2-4} \cmidrule(lr){5-7}
    \text{Caption} & \text{FAD} \(\downarrow\)& \text{IS}\(\uparrow\) & \text{CLAP} \(\uparrow\) & \text{FAD} \(\downarrow\)& \text{IS} \(\uparrow\)& \text{CLAP} \(\uparrow\) \\
    \midrule
    \(\mathcal{D}_o\) & 7.23 & 1.74 & 0.199 & 7.07 & 2.12 & 0.291 \\
    \(\mathcal{D}_g\) & 5.94 & 2.28 & 0.278 & 5.76 & 2.51 & 0.342 \\
    \(\mathcal{D}_\text{merge}\) & \textbf{5.87} & \textbf{2.29} & \textbf{0.284} & \textbf{5.64} & \textbf{2.63} & \textbf{0.350} \\
    \bottomrule
  \end{tabular}
  \caption{Comparison of model performance training on different textual representations, evaluated by FAD, IS and CLAP score.}
\end{table}
\begin{table*}[ht]
  \centering
  
  \scalebox{1.0}{%
    \begin{tabular}{lcc|cccc|cccc}
      \toprule
      & \multicolumn{2}{c}{\text{Details}}
      & \multicolumn{4}{c}{\text{MusicCaps}}
      & \multicolumn{4}{c}{\text{Song Describer Dataset}} \\
      \cmidrule(lr){2-3}\cmidrule(lr){4-7}\cmidrule(lr){8-11}
      \text{Model} & \text{Params} & \text{Hours}
      & \text{FAD} \(\downarrow\) & \text{KL} \(\downarrow\) & \text{IS} \(\uparrow\) & \text{CLAP} \(\uparrow\)
      & \text{FAD} \(\downarrow\) & \text{KL} \(\downarrow\) & \text{IS} \(\uparrow\) & \text{CLAP} \(\uparrow\) \\
      \midrule
      MusicLM               & 1290M & 280k & 4.00 &  –   &  –  &  –  &  –  & – & – & – \\
      MusicGen\textsuperscript{†} & 1.5B  & 20k  & 3.80 & 1.22 &  –  & 0.31 & 5.38 & 1.01 & 1.92 & 0.18 \\
      \midrule
      Mousai                & 1042M & 2.5k & 7.50 & 1.59 &  –  & 0.23 &  –  & – & – & – \\
      Jen-1                 & 746M  & 5.0k & 2.00 & 1.29 &  –  & 0.33 &  –  & – & – & – \\
      AudioLDM 2 – Full     & 712M  & 17.9k& 3.13 & \textbf{1.20} &  –  &  –  &  –  & – & – & – \\
      AudioLDM 2 – Music\textsuperscript{†}
                            & 712M  & 10.8k& 4.04 & 1.46 & 2.67 & 0.34 & 2.77 & 0.84 & 1.91 & 0.28 \\
      \midrule
      Ours (U-Net)          & 1.0B  & 12.5k& 2.03 & 1.51 & 2.41 & 0.33 & \textbf{1.01} & \textbf{0.83} & 1.92 & 0.30 \\
      Ours (QA-MDT)         & \textbf{675M} & 12.5k & \textbf{1.65} & 1.31 & \textbf{2.80} & \textbf{0.35}
                            & 1.04 & \textbf{0.83} & \textbf{1.94} & \textbf{0.32} \\
      \bottomrule
    \end{tabular}%
  }
  
  
  \caption{Objective evaluation results for music generation with diffusion-based
  and language-model-based approaches. Methods we re-inferred are marked with †.}
  \label{tab:main-results}
\end{table*}

\begin{table}[ht]
  \centering
  \label{tab:zhuguan}
  \vspace{1em} 
  
  \fontsize{9}{11}\selectfont 
  \setlength{\tabcolsep}{2pt} 
  
  \begin{tabular}{lcccccccc}
    \toprule
    & \multicolumn{2}{c}{\text{Po}} & \multicolumn{2}{c}{\text{Pmp}} & \multicolumn{2}{c}{\text{Ve}} & \multicolumn{2}{c}{\text{Bg}} \\
    \cmidrule(lr){2-3} \cmidrule(lr){4-5} \cmidrule(lr){6-7} \cmidrule(lr){8-9}
    \text{Model} & \text{Ovl} & \text{Rel} & \text{Ovl} & \text{Rel} & \text{Ovl} & \text{Rel} & \text{Ovl} & \text{Rel} \\
    \midrule
    Ground Truth & 4.00 & 4.00 & 4.47 & 3.60 & 4.10 & 3.80 & 3.87 & 3.87 \\
    \midrule 
    AudioLDM 2 & 2.03 & 2.42 & 3.03 & 3.61 & 3.21 & 3.71 & 3.85 & 3.85 \\
    MusicGen & 2.83 & 3.54 & 2.63 & 2.92 & 3.41 & 3.00 & \textbf{4.33} & 3.83 \\
    Ours(U-Net) & 2.80 & 3.34 & 3.46 & 4.08 & 3.40 & \textbf{3.96} & 3.88 & 3.96 \\
    Ours(QA-MDT) & \textbf{3.27} & \textbf{3.77} & \textbf{3.69} & \textbf{4.19} & \textbf{3.54} & 3.94 & 4.23 & \textbf{4.00} \\
    \bottomrule
  \end{tabular}
  \caption{Evaluation of model performances among different groups, rated for text relevance (\text{Rel}) and overall quality (\text{Ovl}), with higher scores indicating better performance. The groups included Production Operators (\text{Po}), Professional Music Producers (\text{Pmp}), Video Editors (\text{Ve}) and Beginners(Bg)}
\end{table}
We conducted our ablation study on a subset of our training set, which includes ASM and FMA, totaling approximately 3,700 hours and 1.1 million clips. For evaluation, we utilized an out-of-domain set with 1,000 samples randomly selected from MTT~\citep{law2009evaluation}.
Table ~\ref{tab:main-results} compares the model's performance using different textual representations: sentences formed by merging original tags with commas (\(\mathcal{D}_o\)), generated captions (\(\mathcal{D}_g\)), and generated captions refined through filtering and fusion (\(\mathcal{D}_\text{merge}\)). During the filtering and fusion stage, 8.9\% of the generated captions were filtered out, and 15.1\% were fused with original tags using ChatGPT. Each model underwent training for 60,000 steps with a batch size of 64.

  
  
From Table~\ref{tab:main-results} we can also observe consistent trends: employing a captioner to transform audio annotations from sparse words into detailed sentences significantly improved the models’ generalization and diversity. This indicates that detailed annotations are essential for learning the relationship between the models and spectral features. Moreover, the filter and fusion stages led to enhancements across all metrics, highlighting the significance of precise, comprehensive annotations for generalization ability and control ability. We also found that compared to U-Net, the MDT architecture shows stable improvements in basic modeling metrics, making it a better backbone for music spectral modeling.

\subsection{Compared with Previous Methods}
We compared our proposed method with the following representative previous methods: AudioLDM 2~\citep{liu2023audioldm2}, Mousai~\citep{schneider2023mo} and Jen-1~\citep{li2023jen} which model music using spectral latent spaces, MusicLM~\citep{agostinelli2023musiclm}, and MusicGen~\citep{copet2024simple}, which focus on modeling discrete representations.

We re-inferred AudioLDM2-Music and MusicGen-1.5B using their official checkpoints to compare additional metrics under the same environment. The results are presented in Table~\ref{tab:main-results}. For Ours (U-Net), we inferred all text with the prefix ``high quality'', while for Ours (QA-MDT), we used the same prefix along with a \textit{p}-MOS quality token set to level 5. When calculating the CLAP score, we evaluated the generated music with original prompt, which did not include any quality prefix.
The experimental results show significant advantages in both subjective and objective metrics for our models. Since KL divergence measures the distance between audio samples, higher quality audio often results in deviations from the original waveform of Musiccaps, which can lead to lower performance. Although Ours (U-Net) showed a slight FAD advantage on the Song-Describer-Dataset, this may be due to instabilities arising from the small scale test dataset, and we further demonstrated the superiority of QA-MDT in subsequent subjective experiments. Additionally, since MusicGen was trained on non-vocal tracks, it may underperform on captions that include vocals.

Based on subjective evaluation shown in Table~3, our proposed method significantly improves overall audio quality and text-audio consistency, thanks to the label optimization for large music datasets and the quality-aware training strategy. By analyzing the backgrounds of the evaluators and their corresponding results, we can also see that for beginners, the comparison between different systems is not sensitive, which is related to their lack of music background experience and knowledge. However, from the perspective of our method in product operators, video editors, and audio producers, our method offers considerable enhancements, underscoring its potential value to audio industry professionals.
\section{Conclusion and Discussion}\label{sec:dis}

In this study, we address the key challenges in the music generation domain, including model architecture design, large-scale uneven audio quality, and unaligned textual annotations, all of which impede the progress of TTM with quality, musicality, and text alignment.
In the future, we aim to further enhance and expand our model to achieve long-duration, high-sampling-rate, controllable, and highly interactive music generation.
\section*{Acknowledgements}

This work was supported by the National Natural Science Foundation of China under Grant 62171427.
\section*{Statement}

\textbf{Chang Li} and \textbf{Ruoyu Wang} contributed equally.  
\textbf{Jun Du} supervised the entire project and serves as the corresponding author. Further discussion and statement can be found in the Appendix~\ref{sppsta}.

\bibliographystyle{named}

\bibliography{ijcai25}

\appendix
\counterwithin{figure}{section}
\counterwithin{table}{section}

\clearpage  
\section{Appendix}

\subsection{Training Dataset}
\label{sec:databaseaaa}
\begin{table}[h]
  \centering
  \caption{Database statistics. It should be noted that MSD is basically a package of commercial music, which is not a copyright-free content.}
  \vspace{1em} 
  
  \fontsize{9}{11}\selectfont 
  \setlength{\tabcolsep}{2pt} 
  
  \begin{tabular}{lccccc}
    \toprule
    \text{Database} & \text{Duration (h)} & \text{Clip Num} & \text{Sample Rate} & \text{Released} \\
    \midrule
    MTT & 200 & 24K & - & 2009 \\
    FMA & 900 & 10.9K & 44.1kHz & 2016 \\
    MSD & 7333 & 880K & 32kHz & 2011 \\
    ASM & 2777 & 1.0M & mix & 2017 \\
    Pixabay & 1375 & 55K & 44.1kHz & - \\
    \midrule 
    Final & 12.5K & - & 16kHz  & - \\
    \bottomrule
  \end{tabular}
\end{table}

\subsection{Patchify\:Strategy}
In image generation, for latent spaces with consistent aspect ratios, the patchify approach indicates that smaller square patch sizes (\(2 \times 2\) and \(1 \times 1\)) achieve the best results. However, for our audio latent space, which has a size of \(16 \times 256\) with significant aspect ratio imbalance, the traditional square patch pattern may not be the best choice. Therefore, we chose \(2 \times 4\) as the fundamental patch size and conducted comparisons from three aspects: the decoder layer, patchify strategy, and in contrast to the traditional DiT.

Due to the potential inaccuracies inherent in tags, we evaluate our model on the \textit{Generated Captions without Filter and Fusion} to assess its robustness against possibly inaccurate labels training with ASM and FMA, and test on MTT Test Set.
In this section, we aim to explore the fundamental impact of different basic modeling units on spectral. From the experimental comparison of patch sizes \(2 \times 4\), \(1 \times 4\), and \(2 \times 2\) shown in Table \ref{tab:patchsize}, it is evident that reducing the patch size consistently leads to performance improvements due to the more detailed spectral modeling. Considering the inherent spatial correlations in the Fourier transform within music spectrum, we analyze the results of experiments that apply spectral overlaps of 2 and 1 in the time and frequency domains separately, which indicates that introducing overlap in the latent space does indeed contribute to improved results. 
\begin{table}[ht]
\label{table222}
\centering
\caption{Performance comparison between different settings of patchify strategies on MTT Test Set.}
\label{tab:patchsize}
\begin{tabular}{lccccc}
\toprule
\text{Model} & \text{Patch Size} & \text{Overlap Size} & \text{FAD} \(\downarrow\) & \text{KL} \(\downarrow\) & \\
\midrule
DiT & \(2 \times 4\) & \(0 \times 0\) & 6.861 & 4.355\\
MDT & \(2 \times 4\) & \(0 \times 0\) & 5.901 & 3.913 \\
MDT & \(2 \times 2\) & \(0 \times 0\) & 5.685 & 3.820  \\
MDT & \(1 \times 4\) & \(0 \times 0\) & 5.768 & 3.837 \\
MDT & \(2 \times 4\) & \(1 \times 0\) & 5.757 & 3.837 \\
MDT & \(2 \times 4\) & \(0 \times 2\) & \textbf{5.172} & \textbf{3.737}  \\
\bottomrule
\end{tabular}
\end{table}

A smaller patch sizes, such as \(2 \times 1\) or even the more extreme \(1 \times 1\), might offer improved performance. However, due to the significantly increased training and inference costs that accompany these sizes, we refrained from conducting further detailed experiments. Additionally, we explored the impact of these settings in comparison with DiT architectures and discovered that the mask strategy, leveraging the spectral correlations in both the frequency and time domains, markedly enhances spectral modeling, which is evident both in terms of the convergence rate and the quality of the final results. However, based on subjective listening on a small test dataset, we find that although the overlap strategy can significantly improve the model’s objective indicators, it leads to a decline in both melodic generation and aesthetic aspects, which is crucial for music generation. Therefore, we choose \(1 \times 4\) as the base patchify strategy for subsequent experiments and the final model. 

\textbf{Supplementary Note.} We employ patches with a highly unbalanced aspect ratio because, in audio spectrograms, the time axis is typically much longer than the frequency axis and Table A.2 shows no significant performance gain when the temporal granularity is made finer. Furthermore, the perceived decline in melody quality hinges strongly on localized subjective judgments, and it remains unclear whether this phenomenon would differ with larger-scale models and datasets.

\subsection{Comparing with Using ``Negative Prompt''}
\label{sec:negative}

In this section, we aim to compare and analyze the traditional ``negative prompt'' method with our approach. In the implementation of CFG, we use 
\begin{equation}
\small
D_{\theta}^{\text{CFG}}(x, t, y) = D_{\theta}(x, t, y) + w \left(D_{\theta}(x, t, y) - D_{\theta}(x, t, y_{neg})\right),
\end{equation}
where \(y_{neg}\) is formulated as text embedding of "low quality." 

\begin{table}[h]
\centering
\caption{Performance comparison between three systems on MTT-FS.}
\label{tab:system_comp}
\begin{tabular}{lcccc}
\toprule
\text{System} & \text{FAD} \(\downarrow\) & \text{KL} \(\downarrow\) & \text{\textit{p}-MOS} \(\uparrow\) \\
\midrule
MDT  & 5.757 & 3.837  & 3.796\\
MDT + Negative prompt & 5.641 & 3.461 & 3.832 \\
QA-MDT & \textbf{5.200} & \textbf{3.214} & \textbf{4.051} \\
\bottomrule
\end{tabular}
\end{table}

Based on the Table\ref{tab:system_comp}, we found that any form of quality guidance improves the model’s generative performance. However, previous attempts to improve quality relied on the rare instances of "low quality" in the dataset. This necessitated the careful design of numerous negative prompts to avoid generating low-quality results. Furthermore, the text embedding of "low quality" might not be well disentangled during training, leading to an suboptimal results. Furthermore, while alternative guidance strategies (e.g. Autoguidance~\citep{karras2024guiding}) are capable of improving generation quality, they often come at the cost of significantly increased model parameters and sampling steps. In contrast, by incorporating trainable quality awareness, we are able to enhance overall generation quality without sacrificing sampling steps and inference memory consumption. We also leave combining training-based quality-aware strategy and negative-guidance strategy as future work.

Compared with negative-prompt guidance, we believe our approach offers the following \textbf{key advantages}:
\begin{itemize}
    \item As shown in our main experiments, injecting quality information directly at the token-embedding level yields much stronger and more precise control than injecting it through the text encoder’s prompt.
    \item Negative-prompt guidance requires crafting accurate negative descriptions, which increases the burden of prompt engineering and the effort needed to encode specific quality cues, thereby limiting its scalability to other domains and tasks.
    \item When quality attributes are expressed in natural language, they inevitably interact inside the text encoder with other audio-description tokens; this cross-talk between otherwise independent features can lead to sub-optimal outputs.
\end{itemize}

\section*{Supplementary Statement}
\label{sppsta}
\paragraph{Scope of the \textit{p}-MOS Predictor.}
The perceptual MOS (\textit{p}-MOS) scores reported in this study are derived from a music-quality prediction model whose training distribution—owing to the experiment’s temporal constraints—is most closely aligned with vinyl-record characteristics and restricted at 16~kHz. Future work should incorporate a broader, more finely differentiated set of evaluative dimensions that span overall musical quality, musicality, signal fidelity, intrinsic timbre, and general perceptual preference. Expanding both the taxonomic granularity and the pool of human-annotated samples will enable a more accurate, widely applicable quality-assessment framework~\citep{tjandra2025meta}.

\paragraph{Sample Length, Commercial Baselines, and Generality of Our Methods.}
Because open-source academic music datasets are typically short, we trained and evaluated our system on 10-second excerpts and did not benchmark against cutting-edge commercial generators such as \textit{Suno}~\citep{suno2024} or \textit{Udio}~\citep{udio2024}, which can synthesize full-length tracks. Although our U-Net variant naturally scales to arbitrarily long outputs—and diffusion-trajectory fusion can extend generation without retraining~\citep{bar2023multidiffusion,dai2025latent}—we emphasize that our proposed quality-oriented strategies and architectural insights are generation-paradigm agnostic. They can benefit future systems regardless of training regimen, sequence length, channel configuration, sampling rate, or compression constraints. While training on 10-second excerpts provides a practical proxy for validating algorithmic performance, the music modality more readily captures subjective qualities such as melodic appeal and perceived listening quality.

\paragraph{Data Ethics and Annotation Pipeline.}
All training audio was curated to exclude copyright-restricted material, and large-language-model APIs were employed solely for caption refinement under fair-use terms. Nevertheless, more recent multimodal annotators—e.g., \textit{Gemini}~\citep{team2023gemini}, \textit{Kimi-Audio}~\citep{ding2025kimi}, or \textit{Audio Flamingo}~\citep{kong2024audio,ghosh2025audio}—offer richer contextual labeling; integrating them to enhance caption fidelity and diversity is left for future work.

\end{document}